\documentclass[twocolumn,showpacs,amssymb,amsmath,aps,pra]{revtex4-1}
\newcommand{\half}{\mbox{$\frac{1}{2}$}}
\usepackage{graphics}
\usepackage{bm}
\usepackage{graphics}
\usepackage{bm}
\usepackage{amsmath}
\usepackage{amssymb}
\begin{document}
\title{Quantum correlations and least disturbing local measurements}
\author{R. Rossignoli, N. Canosa, L. Ciliberti}
\affiliation{Departamento de F\'{\i}sica-IFLP, Universidad Nacional de La
Plata, C.C. 67, La Plata (1900), Argentina}
\begin{abstract}
We examine the evaluation of the minimum information loss due to an unread
local measurement in mixed states of bipartite systems, for a general entropic
form. Such quantity provides a measure of quantum correlations, reducing for
pure states to the generalized entanglement entropy, while in the case of mixed
states it vanishes just for classically correlated states with respect to the
measured system, as the quantum discord. General stationary conditions are
provided, together with their explicit form for general two-qubit states.
Closed expressions for the minimum information loss as measured by quadratic
and cubic entropies are also derived for general states of two-qubit systems.
As application, we analyze the case of states with maximally mixed marginals,
where a general evaluation is provided, as well as $X$ states and the mixture
of two aligned states.
\end{abstract}
\maketitle
\section{Introduction}
There is currently a great interest on new measures of quantum correlations for
mixed states, different from the entanglement measures \cite{BDSW.96}.  Quantum
entanglement is essential for quantum teleportation \cite{Be.93,NC.00} and also
for pure state quantum computation, where its increase with system size is
necessary to achieve an exponential speedup over classical computation
\cite{JL.03,Vi.03}. However, the computation model proposed by Knill and
Laflamme \cite{KL.98} has shown that for mixed states, such speedup can in
principle be achieved without entanglement \cite{DFC.05}. This suggests the
subsistence of useful quantum correlations in some separable mixed states,
which, we recall, are defined as convex mixtures of product states
\cite{RF.89}. While a separable pure state is a product state, separable mixed
states comprise product states, mixtures of commuting products and also
mixtures of non-commuting product states. The latter can possess entangled
eigenstates and give rise to non-classical capabilities.

Consequently, measures such as the quantum discord
\cite{OZ.01,HV.01,Ve.03,ZZ.03} have recently received much attention. While
coinciding with the entanglement entropy for pure states, the quantum discord
is non-zero for separable mixed states of the last type, vanishing just for
classically or one-way classically correlated states, i.e., states diagonal in
a standard or conditional product basis. The circuit of  \cite{KL.98} was in
fact shown in \cite{DSC.08} to exhibit a non-negligible discord. Other measures
with similar properties include the one-way information deficit
\cite{Ho.05,SKD.11}, the geometric discord \cite{DVB.10}, based on the standard
Hilbert-Schmidt norm, and the general entropic measures which we defined in
\cite{RCC.10}, based on generalized entropic forms. The latter contain the two
previous measures as particular cases, embedding them in a unified picture.
Since they are applicable with entropic forms complying with minimum
requirements, they offer, like the geometric discord, the possibility of easier
evaluations, allowing at the same time to identify some universal features
exhibited by all these measures \cite{RCC.10}. Related generalized measures
vanishing just for full classically correlated states, like those of
\cite{MV.10} and \cite{SL.08}, were also considered \cite{RCC.10}. Let us
remark that important quantum capabilities of separable states with non-zero
discord, and hence non-zero values of the previous measures, were recently
unveiled \cite{MD.11,CAB.11,SKD.11,PGA.11,RRV.11}. Other relevant properties of
the quantum discord and its evaluation in specific states and scenarios were
discussed in \cite{ShL.09,FA.10,FCOC.11,LBAW.08,DG.09,SL2.08,AR.10,SA.09,
WSF.09,CRC.10,AD.11,YL.11}.

The aim of this work is to analyze the explicit evaluation of the generalized
measures of \cite{RCC.10} in some important general cases. We first provide in
Sec.\ \ref{II} the general stationary condition that the least disturbing local
measurement should satisfy, including conditions for its independence from the
entropy employed (universality), together with its explicit form for general
two-qubit states. Here we show that in addition to the quadratic case
(geometric discord), the measure based on a cubic function of the density
matrix (``cubic'' discord) can also be exactly evaluated for any state of two
qubits. Moreover, for two-qubit states this measure shares with the geometric
discord the same pure state limit, where they are both proportional to the
square of the concurrence \cite{WW.98,Ca.03}. As specific applications, we
provide in sec.\ \ref{III} the general expression for two-qubit states with
maximally mixed reduced states, valid for any entropic form, analyzing its main
features. We also examine their evaluation in the so-called $X$ states
\cite{AR.10}, where explicit expressions for the quadratic and cubic cases are
provided, and for the important case of a mixture of two aligned states
\cite{CRC.10}, which represents in particular the exact state of a pair in the
ground state of a finite $XY$ ferromagnetic spin $1/2$ chain in the vicinity of
the factorizing field \cite{RCM.08}. Differences with the quantum discord,
related in particular with the minimizing measurement, are also discussed.
Conclusions are finally drawn in Sec.\ \ref{IV}.

\section{Formalism\label{II}}
\subsection{Information loss by unread local measurement\label{I}}
Let us consider a bipartite system $A+B$ initially in a state $\rho_{AB}$.
After an unread local von Neumann measurement in system $B$, defined by
orthogonal one dimensional  projectors $P_j^B=I\otimes P_j$, with
$P_j=|j_B\rangle\langle j_B|$ ($\sum_j P_j=I$, $P_j P_{j'}=\delta_{jj'}P_j$),
the joint state becomes
\begin{equation}
\rho'_{AB}=\sum_j P_j^B\rho_{AB}P_j^B=\sum_j p_j \rho_{AB/j}\,,
 \label{rp}\end{equation}
where $p_j={\rm Tr}\,\rho_{AB} P_j^B$ is the probability of outcome $j$ and
$\rho_{AB/j}=P_j^B\rho_{AB}P_j^B/p_j$ the state after such outcome. The state
(\ref{rp}) is just the diagonal of $\rho_{AB}$ in a conditional product basis
formed by the states  $|i_j j\rangle\equiv|i_{jA}\rangle|j_B\rangle$, with
$|i_{jA}\rangle$ the eigenstates of $\rho_{A/j}={\rm Tr}_B\rho_{AB/j}$. The
loss of information due to such measurement, i.e., the information contained in
the off-diagonal elements of the original $\rho_{AB}$ in the previous basis,
can be quantified by the quantity \cite{RCC.10}
\begin{equation}
I_f^{M_B}(\rho_{AB})=S_f(\rho'_{AB})-S_f(\rho_{AB})\,,
\label{IfM}\end{equation}
where $S_f(\rho)$ denotes a generalized entropy of the form
\begin{equation}S_f(\rho)={\rm Tr}\,f(\rho)\,,\label{Sf}\end{equation}
with $f:[0,1]\rightarrow\Re$ a smooth strictly concave function ($f''(p)<0$ for
$p\in(0,1)$) satisfying $f(0)=f(1)=0$ \cite{CR.02,WW.78}. This ensures
$S_f(\rho)\geq 0$ for any state $\rho$, with $S_f(\rho)=0$ if and only if
$\rho$ is a pure state ($\rho^2=\rho$), and $S_f(\rho)$ maximum, at fixed
dimension $n$, for the maximally mixed state $\rho=I/n$. Eq.\ (\ref{IfM}) is
then {\it non-negative} for any  $S_f$ of the previous form, vanishing only if
the original $\rho_{AB}$ remains unchanged by such measurement. This positivity
follows from the majorization relation \cite{NC.00,WW.78,Bha.97}
$\rho'_{AB}\prec \rho_{AB}$ ($\rho'_{AB}$ more mixed than $\rho_{AB}$)
satisfied by the post-measurement state, which implies
$S_f(\rho'_{AB})\geq S_f(\rho_{AB})$ for all such $S_f$ \cite{RCC.10}.
Moreover, the previous entropic inequality implies in fact majorization
when valid for {\it all} $S_f$ of the previous form \cite{RC.03}.

The minimum of $I_f^{\rm M_B}$ among all local measurements,
\begin{equation} I_f^B(\rho_{AB})=\mathop{\rm Min}_{M_B}I_f^{M_B}(\rho_{AB})\,,
 \label{If}\end{equation}
provides a measure of the quantum correlations between $A$ and $B$ present in
the original state and destroyed by local measurement \cite{RCC.10}. It
vanishes only if $\rho_{AB}$ is already of the ``classical'' (with respect to
$B$) form (\ref{rp}). For such states there is an unread local measurement in
$B$ ($M_B$) which leaves the state invariant. Eq.\ (\ref{If}) is obviously not
affected by local unitary transformations.

In the case of pure states ($\rho_{AB}^2=\rho_{AB}$), it can be shown that
(\ref{If}) becomes the generalized entanglement entropy
\begin{equation}I_f^{B}(\rho_{AB})=E_f(A,B)\equiv S_f(\rho_A)=S_f(\rho_B),
\end{equation}
where $\rho_A={\rm Tr}_B\,\rho_{AB}$ and $\rho_B$ are the reduced states of
each subsystem \cite{RCC.10}. Hence, pure state entanglement can be seen as the
minimum information loss due to a local measurement. In this case
$I_f^B(\rho_{AB})=I_f^A(\rho_{AB})$, an identity which does not hold in general
for mixed states.

In the von Neumann case $S_f(\rho)=S(\rho)\equiv-{\rm Tr}\rho\log\rho$, Eq.\
(\ref{IfM}) can be also written as \cite{RCC.10}
\begin{equation}
I^{M_B}(\rho_{AB})=S(\rho'_{AB})-S(\rho_{AB})=S(\rho_{AB}||\rho'_{AB})
 \label{IS}\,,\end{equation}
where $S(\rho||\rho')=-{\rm Tr}\,\rho(\log\rho'-\log\rho)$ is the {\it
relative} entropy \cite{NC.00,WW.78,Ve.02} (a non-negative quantity), since
$\rho'_{AB}$ is the diagonal of $\rho_{AB}$ in a certain basis. The minimum
$I^B$ of Eq.\ (\ref{IS}) coincides with the one-way information deficit
\cite{Ho.05,SKD.11} and also with one of the measures discussed in
\cite{MV.10}. In the case of pure states, $I^B$ reduces to the standard
entanglement entropy $E(A,B)=S(\rho_A)=S(\rho_B)$.

In the case of the so-called linear entropy
\begin{equation}S_2(\rho)=2(1-{\rm Tr}\,\rho^2)\,,\label{S2}\end{equation}
which is a quadratic function of $\rho$ and corresponds to
$f(\rho)=2\rho(1-\rho)$ in (\ref{Sf}),  Eq.\ (\ref{IfM}) can be written as
\cite{RCC.10}
\begin{equation}
I_2^{M_B}(\rho_{AB})=S_2(\rho'_{AB})-S_2(\rho_{AB})
 =2||\rho'_{AB}-\rho_{AB}||^2\,,\label{I2}\end{equation}
where $||O||^2={\rm Tr}\,O^\dagger O$ is the squared Hilbert-Schmidt norm. The
ensuing minimum (\ref{If}), to be denoted here as $I_2^B$, becomes then
equivalent \cite{RCC.10} to the geometric discord of ref.\
\cite{DVB.10}, defined as the minimum Hilbert-Schmidt distance between
$\rho_{AB}$ and any classically correlated state of the form (\ref{rp}). In the
case of pure states,  $I_2^B$ reduces to the square of the pure state
concurrence (i.e., the tangle), defined as $C_{AB}=\sqrt{2(1-{\rm
Tr}\,\rho_A^2)}$ \cite{Ca.03}.

In the same way we may define the $q$ information loss
\begin{eqnarray}
I_q^{M_B}(\rho_{AB})&=&S_q(\rho'_{AB})-S_q(\rho_{AB})\,,\label{Iq}\\
S_q(\rho)&=&(1-{\rm Tr}\rho^q)/(1-2^{1-q})\,,\;q>0\,,\label{Sq}
\end{eqnarray}
where $S_q(\rho)$ is the so-called Tsallis entropy \cite{TS.88}, which
corresponds to $f(\rho)=(\rho-\rho^q)/(1-2^{1-q})$ in (\ref{Sf}) and is a
function of the Renyi entropy.  Eq.\ (\ref{Sq}) reduces to the linear entropy
(\ref{S2}) for $q=2$ and to the von Neumann entropy for $q\rightarrow 1$, with
$\log=\log_2$ for the present normalization (chosen such that $S_q(\rho)=1$ for
a maximally mixed single qubit state,  i.e., $2f(1/2)=1$). Eq.\ (\ref{Iq})
allows in particular to switch continuously from the von Neumann case
(\ref{IS}) to the quadratic case (\ref{S2}).

On the other hand, the original quantum discord \cite{OZ.01, HV.01,Ve.03,ZZ.03}
is based on the von Neumann entropy and can be written (considering von Neumann
measurements) as
\begin{equation}
 D^B(\rho_{AB})=\mathop{\rm Min}_{M_B}[I^{M_B}(\rho_{AB})-I^{M_B}(\rho_B)]\,.
 \label{D}\end{equation}
It contains an additional term $I^{M_B}(\rho_B)=S(\rho'_B)-S(\rho_B)$ related
to the local information loss and was actually defined in \cite{OZ.01} as the
minimum difference between the initial mutual information
 \begin{equation} I(A:B)=S(\rho_{A}\otimes\rho_B)-S(\rho_{AB})\,,
 \label{Smut}\end{equation}
where $S(\rho_A\otimes\rho_B)=S(\rho_A)+S(\rho_B)$, and that after the local
measurement, $I^{M_B}(A:B)=S(\rho'_A)+S(\rho'_B)-S(\rho'_{AB})$.
Since $\rho'_A=\rho_A$, such difference reduces to Eq.\ (\ref{D}).

The information loss (\ref{IfM}) can be regarded in fact as a type of
generalized mutual information. Eq.\ (\ref{Smut}) is a measure of the total
correlations between $A$ and $B$ in the original state,  absent in the product
state $\rho_A\otimes\rho_B$. The latter is the state which {\it maximizes} the
von Neumann entropy subject to the constraint of providing just all {\it local}
averages $\langle O\otimes I\rangle$ and $\langle I\otimes O\rangle$, i.e., the
correct reduced states $\rho_{A}$ and $\rho_{B}$. This is in fact what is
expressed by the positivity of Eq.\ (\ref{Smut}): Any other state $\rho_{AB}$
with the same local reduced states has a smaller entropy.

On the other hand, the post-measurement state (\ref{rp}) can be seen as the
{\it more mixed} state providing the same averages as $\rho_{AB}$ of all
observables of the form $\sum_j \alpha_j O_j\otimes P_j$, diagonal in the local
basis defined by $M_B$ (as ${\rm Tr}\,\rho_{AB}\, O\otimes P_j={\rm
Tr}\,\rho'_{AB}\,O\otimes P_j$), such that $S_f(\rho'_{AB})\geq S_f(\rho_{AB})$
$\forall$ $S_f$. The difference $I_f^{M_B}$ is then a measure of the
correlations $\langle O\otimes  |j_B\rangle\langle k_B|\rangle$, $k\neq j$,
contained in the original state $\rho_{AB}$ and absent in $\rho'_{AB}$. In
particular, if $M_B$ is a measurement in a basis where $\rho_B$ is diagonal,
$\rho'_{AB}$ reproduces not only $\rho_A$ ($\rho'_A={\rm
Tr}_B\,\rho'_{AB}=\rho_A$ $\forall$ $M_B$) but also $\rho_B$ ($ \rho'_B={\rm
Tr}_A\,\rho'_{AB}=\rho_B$ for this measurement), as well as all averages
$\langle O\otimes P_j\rangle$, being the more mixed state with such property.
Notice that in contrast with $\rho'_{AB}$, the state $\rho_{A}\otimes \rho_B$
is in general not more mixed than the original state
($\rho_A\otimes\rho_B\prec\!\!\!\!\!/\;\rho_{AB}$), so that the positivity of
Eq.\ (\ref{Smut}) cannot be extended to a general entropy.

\subsection{General stationary condition}

Let us now derive the equations determining the least disturbing local
measurement defined by Eq.\ (\ref{If}). \\
{\it Theorem 1.} For a given entropic function $f$, the least disturbing local
measurement satisfies the equation
\begin{equation}
 {\rm Tr}_A[f'(\rho'_{AB}),\rho_{AB}]=0\label{st}\,,\end{equation}
where $f'$ is the derivative of $f$ and $\rho'_{AB}$ the post-measurement state
(\ref{rp}).  \\
{\it Proof}: The generalized entropy of the state (\ref{rp}) is
\begin{equation}
S_f(\rho'_{AB})=\sum_{i,j}f({p}^i_j)\,,\;\; {p}^i_j=
\langle i_j j|\rho_{AB}|i_j j\rangle\,,\end{equation}
where $\langle i_j j|\rho_{AB}|k_j j\rangle=\delta_{ik} p^i_j$.
Considering a small unitary variation of the local measurement basis, such that
$\delta|j_B\rangle=(e^{i\delta h}-1)|j_B\rangle\approx i\delta h|j_B\rangle$,
with $\delta h$ a small local hermitian operator, we have $\delta
p^i_j\approx i\langle i_{j}j|[\rho_{AB},\delta h_B]|i_{j}j\rangle$ up to first
order in $\delta h$, with $\delta h_B=I\otimes \delta h$.  Hence,
\begin{eqnarray}\delta I_f^{M_B}&=&
\sum_{i,j}f'(p^i_j)\delta p^i_j=i{\rm Tr}\,[f'(\rho'_{AB}),\rho_{AB}]\delta h_B\nonumber\\
&=&i{\rm Tr}_B\,({\rm Tr}_A[f'(\rho'_{AB}),\rho_{AB}])\delta h\,.
\nonumber\end{eqnarray}
The condition $\delta I_f^{M_B}=0$ $\forall$ $\delta h$ leads then to Eq.\ (\ref{st}).

Eq.\ (\ref{st}) implies explicitly $\sum_{i} f'(p^i_j)\langle i_j
j|\rho_{AB}|i_j k\rangle=\sum_i f'(p^i_k)\langle i_k j|\rho_{AB}|i_k k\rangle$
$\forall$ $k,j$, and determines a certain set of feasible local basis
$\{|j_B\rangle\}$. Note that the states $|i_j\rangle$ of $A$ depend in general
on $j$.

The minimizing local basis $\{|j_B\rangle\}$ will not diagonalize, in general,
the reduced state $\rho_B$. Nonetheless, Eq.\ (\ref{st}) entails that the local
eigenstates can be optimum in some important situations: If in a standard
product basis $\{|ij\rangle=|i_A\rangle|j_B\rangle\}$ formed by eigenstates of
$\rho_A$ and $\rho_B$ the only off-diagonal elements of $\rho_{AB}$ are
$\langle ij|\rho_{AB}|kl\rangle$ with $i\neq k$ {\it and} $j\neq l$, such that
\begin{equation}
\langle ij|\rho_{AB}|ik\rangle=\delta_{jk}{p}^i_j\,,\;\;
 \langle ij|\rho_{AB}|lj\rangle=\delta_{il}{p}^i_j \,,
 \label{d}\end{equation}
Eq.\ (\ref{st}) is trivially satisfied {\it $\forall$ $S_f$} for a measurement
in the basis $\{|j_B\rangle\}$. Such basis would then provide  {\it a universal
stationary point of $I_f^B$}. This is precisely the case of a pure state,
written in the Schmidt basis as $|\Psi_{AB}\rangle=\sum_k\sqrt{p_k}|k_A
k_B\rangle$, and also of a mixture of $|\Psi_{AB}\rangle$ with the maximally
mixed state,
\[\rho_{AB}=x|\Psi_{AB}\rangle\langle\Psi_{AB}|+
{\textstyle\frac{1-x}{n}}I\,,\;x\in[0,1]\,,\]
where Eqs.\ (\ref{d}) and hence (\ref{st}) will be satisfied $\forall$ $f$ for a
measurement in the basis $\{|k_B\rangle\}$. It was shown in \cite{RCC.10} that
such basis provides {\it the universal least disturbing local measurement}
for these states, minimizing $I_f^{M_B}$ $\forall$ $S_f$.

In the case of the linear entropy, $f'(\rho'_{AB})\propto I-2\rho'_{AB}$ and
Eq.\ (\ref{st}) becomes just ${\rm Tr}_A[\rho'_{AB},\rho_{AB}]=0$,
indicating that the post-measurement state $\rho'_{AB}$ should locally
(in $B$) commute with the original state.

In the case of the original discord (\ref{D}), the additional local term leads
in the variation to the modified equation
\begin{equation}
{\rm Tr}_A[f'(\rho'_{AB}),\rho_{AB}]
-[f'(\rho'_B),\rho_B]=0\label{stD}\,,\end{equation}
where here $f'(\rho)$ can be replaced by $-\log\rho$.

\subsection{The two-qubit case}
Let us now examine in detail the case of two-qubits.
Any state of a two qubit system can be written as
\begin{eqnarray}\rho_{AB}&=&\frac{1}{4}(I+\bm{r}_A\cdot \bm{\sigma}_A+
\bm{r}_B\cdot \bm{\sigma}_B+\bm{\sigma}_A^t J\bm{\sigma}_B)\,,
 \label{1}\end{eqnarray}
where $\bm{\sigma}_A\equiv \bm{\sigma}\otimes I$, $\bm{\sigma}_B\equiv I\otimes
\bm{\sigma}$, with $\bm{\sigma}^t=(\sigma_x,\sigma_y,\sigma_z)$ the Pauli
operators and $I$ the identity (in the corresponding space).  The basic traces
${\rm tr}\,\sigma_\mu=0$, ${\rm tr}\,\sigma_\mu\sigma_\nu=2\delta_{\mu\nu}$ for
$\mu,\nu=x,y,z$, ensure that
\[\bm{r}_A=\langle \bm{\sigma}_A\rangle\,,\;\bm{r}_B=
\langle \bm{\sigma}_B\rangle\,,\;J=\langle
 \bm{\sigma}_A\bm{\sigma}_B^t\rangle\,,\]
i.e., $J_{\mu\nu}=\langle \sigma_{A\mu}\,\sigma_{B\nu}\rangle$, where $\langle
O\rangle={\rm Tr}\,\rho_{AB}\,O$. 

Any complete local projective measurement in $B$ can be considered as a spin
measurement along the direction of a unit vector $\bm{k}$, represented by the
orthogonal projectors $P_{\pm\bm{k}}=\half(I\pm\bm{k}\cdot\bm{\sigma})$. This
leaves just those elements of $\rho_{AB}$ proportional to
$\bm{k}\cdot\bm{\sigma}$, leading to the post-measurement state
\begin{equation}
\rho'_{AB}
=\frac{1}{4}[I+\bm{r}_A\cdot \bm{\sigma}_A+
(\bm{r}_B\cdot \bm{k})\bm{k}\cdot\bm{\sigma}_B+(\bm{\sigma}_A^t
J\bm{k})(\bm{k}\cdot\bm{\sigma}_B)]\,,\label{rhok}
\end{equation}
which corresponds to $\bm{r}_B\rightarrow \bm{k}\bm{k}^t\bm{r}_B$ and
$J\rightarrow J\bm{k}\bm{k}^t$ in (\ref{1}). The information loss due to this
measurement will be denoted as $I_f^{\bm k}\equiv
S_f(\rho'_{AB})-S_f(\rho_{AB})$.

We now show that {\it the general stationary condition for the measurement
direction $\bm{k}$ in $B$ reads}
\begin{equation}
\alpha_1\bm{r}_B+\alpha_2 J^t\bm{r}_A+\alpha_3 J^tJ\bm{k}=\lambda\bm{k}\,,
 \label{lam}\end{equation}
i.e., $\bm{k}\times(\alpha_1 \bm{r}_B+\alpha_2 J^t\bm{r}_A+\alpha_3
J^tJ\bm{k})=\bm{0}$, where $\lambda$ is a proportionality factor and the
coefficients $\alpha_i$ are given by
\begin{equation}(\alpha_1,\alpha_2,\alpha_3)={\textstyle\frac{1}{4}
\sum\limits_{\mu,\nu=\pm 1}
f'({p}^\mu_\nu)(\nu,\frac{\nu\mu}
{|\bm{r}_A+\nu J\bm{k}|},\frac{\mu}{|\bm{r}_A+\nu J\bm{k}|})}\,,
\label{al}\end{equation}
with $p^\mu_\nu$ ($\mu,\nu=\pm 1$) the eigenvalues of (\ref{rhok}):
\begin{equation}
{p}^\mu_\nu={\textstyle\frac{1}{4}}
(1+\nu \bm{r}_B\cdot\bm{k}+\mu|\bm{r}_A+\nu J\bm{k}|)\,.\label{la}
\end{equation}
{\it Proof:} The state (\ref{rhok}) is diagonal in the conditional product
basis formed by the eigenstates of $\bm{k}\cdot\bm{\sigma}_B$ and
$(\bm{r}_A+\nu J\bm{k})\cdot\bm{\sigma}_A$, with $\nu=\pm 1$ the eigenvalues of
$\bm{k}\cdot\bm{\sigma}_B$, which leads to the eigenvalues (\ref{la}). We can
then write
\[f'(\rho'_{AB})={\textstyle\frac{1}{4}\sum\limits_{\nu,\mu}f'({p}^\mu_\nu)
(I+\mu\frac{\bm{r}_A+\nu J\bm{k}}{|\bm{r}_A+\nu
 J\bm{k}|}\cdot\bm{\sigma}_A)(I+\nu\bm{k}\cdot\bm{\sigma}_B)}\,.\]
Using now the basic trace relations and
$[\bm{r}\cdot\bm{\sigma},\bm{s}\cdot\bm{\sigma}]=
2i(\bm{r}\times\bm{s})\cdot\bm{\sigma}$, we obtain ${\rm
Tr}_A\,[(\bm{r}\cdot\bm{\sigma}_A)
(\bm{s}\cdot\bm{\sigma}_B),\bm{\sigma}_A^tJ\bm{\sigma}_B]= 4i(\bm{s}\times
J^t\bm{r})\cdot\bm{\sigma}_B$ and hence
\[{\rm Tr}_A\,[f'(\rho'_{AB}),\rho_{AB}]={\textstyle}i[\bm{k}\times(\alpha_1\bm{r}_B
 +\alpha_2 J^t\bm{r}_A+\alpha_3 J^tJ\bm{k})]\cdot\bm{\sigma}_B\,,\]
with $\alpha_i$ given by (\ref{al}). Eq.\ (\ref{st}) leads then to Eq.\
(\ref{lam}).

We can also check Eq.\ (\ref{lam}) directly. From (\ref{la}), we have $\delta
{p}^\mu_\nu= \frac{\nu}{4}(\bm{r}_B+\mu\frac{J^t(\bm{r}_A+\nu
J\bm{k})}{|\bm{r}_A+\nu J\bm{k}|})\cdot\delta\bm{k}$ for changes $\delta\bm{k}$
in the direction of the local measurement apparatus, with
$\bm{k}\cdot\delta\bm{k}=0$ since $\bm{k}$ is a unit vector. The condition
$\delta I_f^{\bm k}=\sum_{\nu,\mu}f'({p}^\mu_\nu)\delta {p}^\mu_\nu=0$ then
implies $(\alpha_1\bm{r}_B+\alpha_2 J^t\bm{r}_A+\alpha_3
J^tJ\bm{k})\cdot\delta\bm{k}=0$, which leads to Eq.\ (\ref{lam}) since
$\delta\bm{k}$ is orthogonal to $\bm{k}$.

Writing $\bm{k}=(\sin\gamma\cos\phi,\sin\gamma\sin\phi,\cos\gamma)$, Eq.\
(\ref{lam}) leads to a transcendental system for $\gamma,\phi$
($\tan\gamma=d_z/\sqrt{d_x^2+d_y^2}$, $\tan\phi=d_y/d_x$, with $\bm{d}$ the
l.h.s.\ of (\ref{lam})). Eq.\ (\ref{lam}) can be also seen as a self-consistent
eigenvalue equation for the matrix $(\alpha_1\bm{r}_B+\alpha_2
J^t\bm{r}_A)\bm{k}^t+\alpha_3 J^tJ$.

Let us remark that the initial reduced local state $\rho_B={\rm Tr}_A\,
\rho_{AB}={\textstyle\frac{1}{2}} (I+\bm{r}_B\cdot\bm{\sigma})$, becomes
 \begin{equation}
\rho'_B={\textstyle\frac{1}{2}}[I+(\bm{r}_B\cdot\bm{k})
(\bm{k}\cdot\bm{\sigma})]\,,
 \label{rbk}\end{equation}
after the local measurement. The minimizing direction $\bm{k}$ will depend on
the matrix $J$ and may obviously deviate from $\bm{r}_B$, changing the local
state. A ``transition'' in the direction of the least disturbing $\bm{k}$, from
$\bm{r}_B$ to the direction of the main eigenvector of $J^tJ$, can then be
expected from (\ref{lam}) as $J$ increases from $0$, whose details will in
general depend on the choice of entropy (see sec.\ \ref{III}).

In the case of the original quantum discord (\ref{D}), the extra local
contribution in (\ref{stD}) leads to the modified stationary condition
(see also \cite{AD.11})
\begin{eqnarray}
(\alpha_1-\eta)\bm{r}_B+\alpha_2 J^t\bm{r}_A+\alpha_3 J^tJ\bm{k}
=\lambda\bm{k}\,,\label{eqmod}
\end{eqnarray}
where $\eta=\frac{1}{2}\sum_{\nu=\pm}\nu f'(p_\nu)=\half\log(p_-/p_+)$, with
$p_\nu=\sum_\mu {p}^\mu_\nu=\frac{1}{2}(1+\nu\bm r_B\cdot\bm k)$
the eigenvalues of $\rho'_B$. 
The extra term $-\eta\bm{r}_B$ will tend to
diminish the effect of $\bm{r}_B$, favoring the direction determined  by
$J^tJ$.

\subsection{The quadratic and cubic information measures}
While the evaluation of a general entropy $S_f(\rho)$ requires the
determination of the eigenvalues of $\rho$, for those choices of $f$ involving just
low integer powers of $\rho$, $S_f(\rho)$ can be determined without their
explicit knowledge. For instance, using just the basic trace relations
${\rm tr}\,\sigma_\mu=0$. ${\rm tr}\,\sigma_\mu\sigma_\nu=2
\delta_{\mu\nu}$,
the linear entropy (\ref{S2}) of any two qubit state can be evaluated as
\begin{equation}
S_2(\rho_{AB})=
{\textstyle\frac{3}{2}}-{\textstyle\frac{1}{2}}(|\bm{r}_A|^2+|\bm{r}_B|^2+||J||^2)\,,
\label{S22}\end{equation}
where $||J||^2={\rm tr}\,J^tJ$ and $|\bm{r}|^2=\bm{r}\cdot\bm{r}=\bm{r}^t\bm{r}$.
For the post-measurement state (\ref{rhok}), Eq.\ (\ref{S22}) becomes
 \begin{eqnarray}S_2(\rho_{AB}')&=&
{\textstyle\frac{3}{2}}-
{\textstyle\frac{1}{2}}|\bm{r}_A|^2
-{\textstyle\frac{1}{2}}\bm{k}^tM_2\bm{k},\label{S22p}\\
 M_2&=&\bm{r}_B\bm{r}_B^t+J^tJ\,, \label{M2}\end{eqnarray}
where $M_2$ is a positive semidefinite symmetric matrix.

The information loss becomes therefore
\begin{equation}
I_2^{\bm{k}}={\textstyle\frac{1}{2}}(|\bm{r}_B|^2+||J||^2 -\bm{k}^tM_2\bm{k})=
 {\textstyle\frac{1}{2}}({\rm tr}\,M_2-\bm{k}^tM_2\bm{k})\,.
\end{equation}
The minimum $I_2^{\bm{k}}$ is just twice the {\it geometric discord}, defined
and evaluated for two qubits in \cite{DVB.10}. It corresponds then to $\bm{k}$
directed along the eigenvector with  the {\it largest} eigenvalue of the matrix
$M_2$:
\begin{eqnarray}
I_2^B(\rho_{AB})&=&\mathop{\rm Min}_{\bm k}\,
I_2^{\bm k}={\textstyle\frac{1}{2}}({\rm tr}\,M_2-\lambda_1)
={\textstyle\frac{1}{2}}(\lambda_2+\lambda_3)
 \label{I22}\end{eqnarray}
where $(\lambda_1,\lambda_2,\lambda_3)$ are the eigenvalues of $M_2$ sorted in
{\it decreasing} order. A state $\rho_{AB}$ which is already of the form
(\ref{rhok}) leads to $I_f^B(\rho_{AB})=0$ $\forall$ $S_f$ and is then
characterized by a matrix $M_2$ of rank 1 (such that $\lambda_2=\lambda_3=0$).
It is verified that for $f'({p}^\mu_\nu)\propto 1-2{p}^\mu_\nu$, Eq.\
(\ref{lam}) reduces to the present eigenvalue equation $M_2\bm{k}=\lambda
\bm{k}$, since $(\alpha_1,\alpha_2,\alpha_3)\propto (\bm{r}_B\cdot\bm{k},0,1)$.

Another entropy which can be easily evaluated for any state
of two qubits is the $q=3$ case in (\ref{Sq}),
\begin{equation}
 S_3(\rho)={\textstyle\frac{4}{3}}(1-{\rm Tr}\,\rho^3)\,.
 \label{S3}\end{equation}
{\it Theorem 2}. The entropy (\ref{S3}) of the general two qubit state (\ref{1}), 
and the ensuing minimum information loss $I_3^B(\rho_{AB})$ 
due to a local measurement in $B$, are given by 
\begin{eqnarray}
S_3(\rho_{AB})
&=&{\textstyle\frac{1}{2}}[S_2(\rho_{AB})+1-(\bm{r}_A^t\,J\,\bm{r}_B-{\rm det}\,J)]
\,,\label{S33}\\
I_3^B(\rho_{AB})&=&\mathop{\rm Min}_{\bm k}\,I_3^{\bm k}
={\textstyle\frac{1}{4}}({\rm tr}\,M_3-2\,{\rm det}\,J-\lambda_1)\nonumber\\
&=&{\textstyle\frac{1}{4}}(\lambda_2+\lambda_3)- {\textstyle\frac{1}{2}}\, {\rm
det}\,J \label{I33}\,,
\end{eqnarray}
where $S_2(\rho_{AB})$ is the entropy (\ref{S22}) and $(\lambda_1,\lambda_2,\lambda_3)$
are the eigenvalues, sorted in decreasing order, of the matrix
\begin{equation}
M_3=\bm{r}_B\bm{r}_B^t+J^tJ+\bm{r}_B\bm{r}_A^tJ+J^t\bm{r}_A\bm{r}_B^t\,,
 \label{M3}\end{equation}
 which is positive semidefinite. \\
 {\it Proof}: Applying the basic trace relations together with  ${\rm
tr}\,\sigma_\mu\sigma_\nu \sigma_\tau=2i\epsilon_{\mu\nu\tau}$, with $\epsilon$
the full antisymmetric tensor ($\mu,\nu,\tau\in\{x,y,z\}$), the only terms 
with non-zero trace in $\rho^3$ are ${\rm Tr}(\bm{r}_A^t{\sigma}_A)(\bm{\sigma}_A^tJ\bm{\sigma_B})
(\bm{r}_B^t\bm{\sigma}_B)=4\bm{r}_A^tJ\bm{r}_B$ (and the same for its 3!
permutations), ${\rm Tr}(\bm{\sigma}_A^t J \bm{\sigma}_B)^3=3!(2i)^2 {\rm
det}\,J$ and the quadratic terms appearing already in ${\rm Tr}\,\rho^2$. This leads to 
to Eq.\ (\ref{S33}).

Using Eq.\ (\ref{S33}), the cubic entropy of the post-measurement
state (\ref{rhok}) can  be expressed as
\begin{equation}
S_3(\rho'_{AB})={\textstyle\frac{5}{4}}-{\textstyle\frac{1}{4}}(|\bm{r}_A|^2
 +\bm{k}^tM_3\bm{k})\,,\label{S33p}\end{equation}
where $M_3$ is the matrix (\ref{M3}), since $\bm{r}_A^tJ\bm{r}_B={\rm
tr}\,\bm{r}_B\bm{r}_A^tJ={\rm tr}\,J^t\bm{r}_A\bm{r}_B^t$ and ${\rm det}
(J\bm{k}\bm{k}^t)=0$.  The matrix $M_3$ is clearly symmetric and also {positive
semi-definite}, as $\bm{k}^t M_3\bm{k}\geq
(|\bm{k}\cdot\bm{r}_B|-|J\bm{k}|)^2\geq 0$ $\forall$ $\bm{k}$ if
$|\bm{r}_A|\leq 1$.  The information loss $I_3^{\bm
k}=S_3(\rho'_{AB})-S_3(\rho_{AB})$ is therefore
\begin{equation}
I_3^{\bm{k}}={\textstyle\frac{1}{4}}({\rm tr}\,M_3-2\, {\rm det}\,J-\bm{k}^t
 M_3\bm{k})\,,\end{equation}
where ${\rm tr}\,M_3=|\bm r_B|^2+||J||^2+2\bm{r}_A^tJ\bm{r}_B$. Its minimum
corresponds then to $\bm{k}$ along the eigenvector with the {\it largest
eigenvalue of $M_3$}, which leads to Eq.\ (\ref{I33}).

It is also verified that Eq.\ (\ref{lam}) leads in the present case to the same
eigenvalue equation $M_3\bm{k}=\lambda\bm{k}$, since
$(\alpha_1,\alpha_2,\alpha_3)
\propto(\bm{r}_B^t\bm{k}+\bm{r}_A^tJ\bm{k},\bm{r}_B^t\bm{k},1)$ for
$f'(p^\mu_\nu)\propto 1-3(p^\mu_\nu)^2$. As opposed to $I_2^{\bm{k}}$, the
minimizing measurement can now depend also on $\bm{r}_A$ through the last terms
of $M_3$. A state of the form (\ref{rhok}) is  then characterized by matrices
$M_3$ and $J$ of rank $1$, such that Eq.\ (\ref{I33}) vanishes.

Let us notice that under arbitrary local rotations
$\bm{\sigma}_\alpha\rightarrow R_\alpha\bm{\sigma}_\alpha$ for $\alpha=A,B$
($R_\alpha R_\alpha^t=I$, ${\rm det}\,R_\alpha=+1$), we have
$\bm{r}_\alpha\rightarrow R_\alpha^t\bm{r}_\alpha$ and $J\rightarrow R_A^t J
R_B$ in (\ref{1}), such that $M_2\rightarrow R_B^t M_2R_B$ and $M_3\rightarrow
R_B^t M_3R_B$. Their eigenvalues remain therefore invariant. Of course, ${\rm
det}\,J$ and all other terms in  Eqs.\ (\ref{S22}) and (\ref{S33}) remain also
unaltered.

Eqs.\ (\ref{S22}) and (\ref{S33}) provide in fact strict bounds on these
invariants. As $S_2(\rho_{AB})\geq 0$ $\forall$ $\rho_{AB}$,  Eq.\ (\ref{S22})
implies
\begin{equation}|\bm{r}_A|^2+|\bm{r}_B|^2+||J||^2\leq 3\,,\label{b1}\end{equation}
with $|\bm{r}_A|^2+|\bm{r}_B|^2+||J||^2=3$ if and only if $\rho_{AB}$ is pure
($\rho_{AB}^2=\rho_{AB}$, $S_2(\rho_{AB})=0$). Moreover, as ${\rm Tr}\,
\rho^{q'}\leq {\rm Tr}\,\rho^q$ if $q'>q>0$, for the present normalization we
have $S_3(\rho)\geq \frac{2}{3}S_2(\rho)$, which for a two qubit state implies
\begin{equation}
\bm{r}_A^tJ\bm{r}_B-{\rm det}\,J\leq 1-{\textstyle\frac{1}{3}}S_2(\rho_{AB})\,,
 \label{b2}\end{equation}
with $\bm{r}_A^tJ\bm{r}_B-{\rm det}\,J=1$ if and only if $\rho_{AB}$ is pure.
We can verify these results by writing a pure state of two qubits in the
Schmidt basis, $|\Psi_{AB}\rangle=\sqrt{p}\,|00\rangle+\sqrt{1-p}\,|11\rangle$,
with $p\in[0,1]$, which leads to $|\bm{r}_A|=|\bm{r}_B|=|2p-1|$,
$||J||^2=1+8p(1-p)$, $\bm{r}_A^tJ\bm{r}_B=(2p-1)^2$ and ${\rm
det}\,J=-4p(1-p)$, and hence to equality in (\ref{b1})--(\ref{b2}).

An important final remark concerning the quadratic and cubic entropies is that
for an arbitrary single qubit state
$\rho_A=\frac{1}{2}(I_2+\bm{r}_A\cdot\bm{\sigma})$ they are {\it identical},
since ${\rm tr}\sigma_\mu^m=0$ for $m$ odd:
\begin{equation}S_2(\rho_A)=S_3(\rho_A)=1-|\bm{r}_A|^2\,.\label{23}\end{equation}
This entails that the corresponding entanglement monotones \cite{Vi.00} for a
two-qubit state are also {\it identical} \cite{RCC.10}, coinciding with the
square of the concurrence $C_{AB}$ \cite{WW.98,Ca.03}.
{\it Both } quantities $I_2^B$ and $I_3^B$ reduce then to the squared
concurrence $C^2_{AB}$ in the case of a pure two-qubit state.

This last result can be directly verified using the previous Schmidt
decomposition: Both matrices $M_2$ and $M_3$ become diagonal in the ensuing $z$
basis, their two lowest eigenvalues being identical:
$\lambda_2=\lambda_3=4p(1-p)=-{\rm det}\,J$. Eqs.\ (\ref{I22}) and (\ref{I33})
lead then to $I_2^B=I_3^B=4p(1-p)$, which is just the square of
$C_{AB}=2\sqrt{p(1-p)}$.

\section{Application\label{III}}
\subsection{States with maximally mixed reduced states}
As a first example, let us consider the case $\bm{r}_A=\bm{r}_B=\bm{0}$ in
(\ref{1}), such that  $\rho_A=\rho_B=\frac{1}{2}I$ and
\begin{equation}
\rho_{AB}={\textstyle\frac{1}{4}}
 (I+\bm{\sigma}_A^tJ\bm{\sigma}_B)\,.
 \label{rhox}\end{equation}
We will show that for the state (\ref{rhox}): \\
a) The measurement direction $\bm{k}$ in system $B$  minimizing $I_f^B$ is {\it
universal}, i.e., the same for any entropy $S_f$, and given by that of the eigenvector
with the largest eigenvalue of the matrix $J^tJ$. \\
b) The ensuing minimum information loss is given by
\begin{eqnarray}
I_f^B(\rho_{AB})&=&{\textstyle 2f(\frac{p_1+p_2}{2})+2f(\frac{p_3+p_4}{2})}\nonumber\\
 &&-f(p_1)-f(p_2)-f(p_3)-f(p_4)\,, \label{ff}\end{eqnarray}
where $(p_1,p_2,p_3,p_4)$ are the eigenvalues of (\ref{rhox}) {\it
sorted in  decreasing order}. \\
c) $I_f^A=I_f^B$ $\forall$ $f$, the minimizing direction in $A$ being that of
the eigenvector with the largest eigenvalue of $JJ^t$. \\
{\it Proof of a):} For $\bm{r}_A=\bm{r}_B=\bm{0}$, the eigenvalues (\ref{la})
of $\rho'_{AB}$ become ${p}^\mu_\nu(\bm{k})=\frac{1}{4}(1+\nu\mu |J\bm{k}|)$,
being two-fold degenerate. If $\bm{k}_m$ is the normalized eigenvector with the
largest eigenvalue ($J_m^2$) of $J^tJ$, we have
$|J\bm{k}|=\sqrt{\bm{k}^tJ^tJ\bm{k}}\leq \sqrt{\bm{k}^t_m J^tJ\bm{k}_m}=|J_m|$
for any unit vector $\bm{k}$, and hence $p^\mu_\mu(\bm{k})\leq
p^\mu_\mu(\bm{k}_m)$. This implies that the distribution
$\{{p}^\nu_\mu(\bm{k})\}$ is {\it majorized} \cite{Bha.97} by
$\{{p}^\nu_\mu(\bm{k}_m)\}$,  i.e.,
\begin{equation}
\rho'_{AB}(\bm{k})\prec \rho' _{AB}(\bm{k}_m)=
{\textstyle\frac{1}{4}}[I+J_m(\tilde{\bm{k}}_m\cdot
 \sigma_{A})(\bm{k}_m\cdot\sigma_{B})] \,,\label{rkm}\end{equation}
where $\tilde{\bm{k}}_m=J\bm{k}_m/J_m$ is the corresponding eigenvector of
$JJ^t$, entailing $S_f(\rho'_{AB}(\bm{k}))\geq S_f(\rho'_{AB}(\bm{k}_m))$ and
hence $I^{\bm {k}}_f\geq I_f^{{\bm k}_m}$ $\forall$ $\bm{k}$ and $S_f$. The
state $\rho' _{AB}(\bm{k}_m)$ is thus the {\it least mixed} classical state
associated with $\rho_{AB}$, and measurement along $\bm{k}_m$ the {\it least
disturbing local measurement} (in $B$) for {\it any} $S_f$. Accordingly, the
general stationary condition (\ref{lam}) leads in this case to the eigenvalue
equation $J^tJ\bm{k}=\lambda\bm{k}$ $\forall$ $f$, with both matrices $M_2$ and
$M_3$ of  Eqs.\ (\ref{M2}), (\ref{M3}) reducing to $J^tJ$.

This result is apparent. The local axes can be always chosen such that the
matrix $J$ is {\it diagonal}. This can be achieved through its singular value
decomposition $J=U_AJ^dU_B^t$, where $J^d_{\mu\nu}=J_\mu\delta_{\mu\nu}$, with
$J_\mu^2$ the eigenvalues of  $J^tJ$ (the same as those of $JJ^t$) and $U_A$,
$U_B$ orthonormal  matrices ($U_\alpha U^t_\alpha=I$). The signs of the $J_\mu$
should be chosen such that $U_\alpha$ are rotation matrices (${\rm
det}\,U_\alpha=+1$). Replacing $\bm{\sigma}_\alpha\rightarrow
U_\alpha\bm{\sigma}_\alpha$ in (\ref{rhox}), we then obtain
 \begin{equation}\rho_{AB}=
 {\textstyle\frac{1}{4}}(I+\sum_{\mu=x,y,z}J_\mu \sigma_{A\mu}\sigma_{B\mu})
 \,.\label{rhoxd}\end{equation}
Since $|J_m|={\rm Max}\{|J_\mu|\}$, the universal least disturbing measurement
is, therefore, {\it along the maximally correlated direction}, leaving the
largest term of (\ref{rhoxd}) in the post-measurement state (\ref{rkm}). Note
that Eq.\ (\ref{rhoxd}) satisfies Eqs.\ (\ref{d}) in a product basis formed by
the eigenstates of $\sigma_{A\mu}\sigma_{B\mu}$, for any $\mu=x,y,z$. \\
{\it Proof of b):} Eq.\ (\ref{rhoxd}) is diagonal in the Bell basis
$\{|\Psi_{1,2}\rangle=\frac{|00\rangle\pm|11\rangle}{\sqrt{2}}$,
$|\Psi_{3,4}\rangle=\frac{|01\rangle\pm|10\rangle}{\sqrt{2}}\}$, i.e.,
$\rho_{AB}=\sum_{i}p_i|\Psi_i\rangle\langle \Psi_i|$, with eigenvalues
\[{\textstyle p_{1,2}=\frac{1+J_z\pm (J_x-J_y)}{4},\;
 p_{3,4}=\frac{1-J_z\pm (J_x+J_y)}{4}}\,.\]
Without loss of generality we may always choose the local axes $x,y,z$ such
that $|J_m|=|J_z|\geq |J_x|\geq |J_y|$, with $J_z\geq 0$, $J_x\geq 0$
(rotations of angle $\pi$ around one of the axes in $A$ or $B$ lead to
$J_\mu\rightarrow-J_\mu$ for the other axes). In such a case $p_1\geq p_2\geq
p_3\geq p_4$, and the least disturbing measurement is along $z$, such that Eq.\
(\ref{rkm}) becomes
\begin{equation}
\rho'_{AB}(\bm{k}_m)={\textstyle\frac{1}{4}}(I+J_z\sigma_{A z}\sigma_{B z})
 \,,\end{equation}
having degenerate eigenvalues
\[{\textstyle \frac{1+J_z}{4}=\frac{p_1+p_2}{2}\,,\;\;
 \frac{1-J_z}{4}=\frac{p_3+p_4}{2}}\,.\]
The minimum information loss $I^B_f(\rho_{AB})=
S_f(\rho'_{AB}(\bm{k}_m))-S_f(\rho_{AB})$ becomes therefore Eq.\ (\ref{ff}),
where $(p_1,p_2,p_3,p_4)$ are in general the eigenvalues of $\rho_{AB}$
sorted in  decreasing order. \\
{\it Proof of c):} Since Eq.\ (\ref{ff}) is fully determined by the sorted
eigenvalues of $\rho_{AB}$, we have obviously $I_f^A=I_f^B$, a result which is
apparent from the symmetric representation (\ref{rhoxd}). From (\ref{rkm}) it
is seen that the minimizing measurement in $A$ is along $\tilde{\bm{k}}_m$.

Let us now discuss the main features of Eq.\ (\ref{ff}). It is verified that
the strict concavity of $f$ ensures $I_f^B(\rho_{AB})\geq 0$ $\forall$ $S_f$,
with $I_f^B(\rho_{AB})=0$ {\it only if}  $p_1=p_2$ {\it and} $p_3=p_4$, in
which case $\rho_{AB}=\rho'_{AB}=p_1(|00\rangle\langle 00|+|11\rangle\langle
11|)+p_3(|01\rangle\langle 01|+|10\rangle\langle 10|)$  {\it is a classically
correlated state}.

In the von Neumann case $f(p)=-p\log p$, Eq.\ (\ref{ff}) is just the quantum
discord $D^A=D^B$ of the state, coinciding with the result of ref.\
\cite{SL2.08}. For the states (\ref{rhox}), $\rho'_B=\rho_B=\half I$ for any
$M_B$, entailing that the quantum discord (\ref{D}) reduces to the information
deficit, i.e., to the present quantity $I_f^B$ for the von Neumann choice of
$f$.

In the quadratic case (\ref{S2}),  Eq.\ (\ref{I22}) or (\ref{ff}) lead to
\begin{equation}
I_2^B(\rho_{AB})={\textstyle\frac{1}{2}}(J_x^2+J_y^2)=(p_1-p_2)^2+(p_3-p_4)^2\,,
\label{I2s}
\end{equation}
which is just twice the geometric discord of the state,
whereas in the cubic case (\ref{S33}), Eqs.\ (\ref{I33}) or (\ref{ff}) lead to
\begin{eqnarray}
I_3^B(\rho_{AB})&=&{\textstyle\frac{1}{4}
(J_x^2+J_y^2)-\frac{1}{2}J_xJ_yJ_z}\label{I3sa}\\
&=&(p_1-p_2)^2(p_1+p_2)+(p_3-p_4)^2(p_3+p_4)\label{I3s}
\end{eqnarray}
which is just the average of the terms in (\ref{I2s}) and implies
$I^B_3(\rho_{AB})\leq I^B_2(\rho_{AB})$.

Let us notice that for small $J_\mu$, Eq.\ (\ref{ff}) becomes in fact
proportional to (\ref{I2s}) {\it for any} $S_f$: Setting $J_m=J_z$,
\begin{equation}
I_f^B(\rho_{AB})\approx {\textstyle\frac{1}{2}}
c_f (J_x^2+J_y^2)+O(J^3)=c_f I_2^B(\rho_{AB})+O(J^3)
\end{equation}
with $c_f=-\frac{1}{4}f''(\frac{1}{4})>0$. This implies {\it a universal
behavior} in the vicinity of the maximally mixed state $I/4$, in agreement with
the general results of \cite{RCC.10}.

{\it Relation with entanglement}. It is well known that the state (\ref{rhox})
is entangled if and only if its largest eigenvalue $p_1$ satisfies $p_1>1/2$.
Its concurrence \cite{WW.98} is given by
\begin{equation}
C_{AB}={\rm Max}[2p_1-1,0]\,,\label{C}
\end{equation}
with $2p_1-1=p_1-p_2-p_3-p_4$. This implies
\begin{equation}
 I_2^B\geq C^2_{AB},\;\;I_3^B\geq C^2_{AB}\,,\label{bound}
 \end{equation}
with equality for $C_{AB}>0$ valid in both cases only if $p_3=p_4=0$
($C^2_{AB}\leq (p_1-p_2)^2-(p_1-p_2)(p_3+p_4)\leq (p_1-p_2)^2(p_1+p_2)$ if
$p_3+p_4\leq p_1-p_2$). Eq.\ (\ref{bound}) means that for the states
(\ref{rhox}), $I_2^B$ and $I_3^B$ are both {\it upper bounds} to their
corresponding entanglement monotone. This is not a general property. For
instance, it is not valid in the von Neumann case $f(\rho)=-\rho\log\rho$,
where Eq.\ (\ref{ff}) can be lower than the entanglement of formation
$E_{AB}=\sum_{\nu=\pm}f(\frac{1+\nu\sqrt{1-C_{AB}^2}}{2})$ \cite{WW.98} for the
present states.

\begin{figure}
\vspace*{-0.cm}

\centerline{\scalebox{.7}{\includegraphics{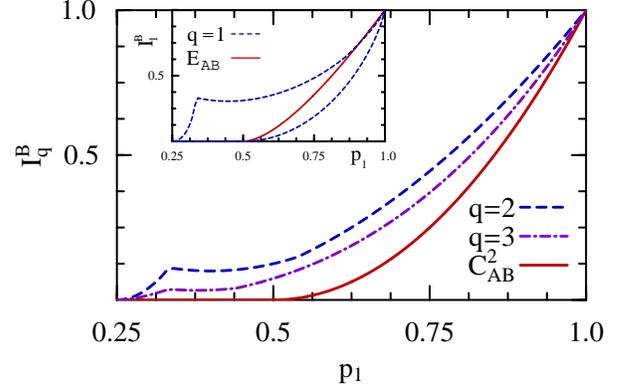}}}
 \vspace*{-0.5cm}

\caption{The maximum and minimum values reached by the quantum
correlation measures $I_2^B(\rho_{AB})$ and $I_3^B(\rho_{AB})$ in the state
(\ref{rhox}), Eqs.\ (\ref{I2s})--(\ref{I3s}), as a function of its maximum
eigenvalue $p_1$. The common minimum is just the squared concurrence
$C^2_{AB}$, whereas the respective maximum is indicated by the dashed and
dashed-dotted lines. The inset depicts the maximum and minimum values reached
in this state by $I_f^B$ in the von Neumann case ($q=1$, $log=log_2$), where it coincides
with the quantum discord, with the solid line depicting the entanglement of
formation. The least disturbing measurement is here the same for all entropies,
and along the direction of the main principal axis of $J^tJ$ (see text).
Quantities plotted are dimensionless in all figures.}
 \label{f1}
\end{figure}
 \vspace*{0.cm}

Fig.\ \ref{f1} depicts the maximum and minimum values reached by $I_2^B$ and
$I_3^B$ in the states (\ref{rhox}) for fixed values of the maximum eigenvalue
$p_1$. The common minimum is just the squared concurrence $C^2_{AB}$, reached
for $p_3=p_4=0$ if $p_1\geq 1/2$ (and $p_2=p_1$, $p_3=p_4$ if $p_1\leq 1/2$).
The maximum is reached for $p_2=p_3=p_4$ if $p_1\geq 7/13\approx 0.54$ for
$I_2$ and $p_1\agt 0.44$ for $I_3$, and for $p_2=p_3$, $p_4=0$, if  $p_1$ lies
below the previous values and above $1/3$. As a result, the maximum values for
zero concurrence of $I_2$ and $I_3$ within these states are $1/8$ and $2/27$
respectively, obtained at $p_1=1/2$.

In contrast, in the von Neumann case the minimum (again obtained for
$p_3=p_4=0$ if $p_1\geq 1/2$) lies clearly below $E_{AB}$ $\forall$
$p\in(1/2,1)$, and even the maximum (attained at $p_2=p_3=p_4$ if $p\agt 0.86$
and $p_2=p_3$, $p_4=0$ if  $1/3\leq p_1\alt 0.86$) lies {\it below} $E_{AB}$ if
$p_1\agt 0.91$. If $p\leq 1/3$ the maximum in these three measures is reached
for $p_2=p_3=p_1$.

\subsection{States with parity symmetry}
Let us now consider the case where both $\bm{r}_A$ and $\bm{r}_B$ are directed
along the same principal axis, i.e., $\bm{r}_B$ along $\bm{k}$ and $\bm{r}_A$
along $J\bm{k}$, with $\bm{k}$ an eigenvector of $J^tJ$ (and hence, $J\bm{k}$
an eigenvector of $JJ^t$). Choosing these axes as the local $z$ axes, such that
$\bm{r}_A=r_A\bm{k}_z$, $\bm{r}_B=r_B\bm{k}_z$ and
$J_{\mu\nu}=J_\mu\delta_{\mu\nu}$, such state can be written as
\begin{eqnarray}\rho_{AB}&=&{\textstyle\frac{1}{4}}(I+r_A\sigma_{Az}+r_B\sigma_{Bz}+
\sum_{\mu=x,y,z}J_\mu \sigma_{A\mu}\sigma_{B\mu})\label{X}\\
&=&\frac{1}{4}\left(\begin{array}{cccc}a_+&0&0&\alpha_+\\0&c_+&\alpha_-&0\\
0&\alpha_-& c_-&0\\ \alpha_+&0&0&a_- \end{array}\right),
\begin{array}{rcl}
a_{\pm}&=&1+J_z\pm (r_A+r_B)\\ c_{\pm}&=&1-J_z\pm(r_A-r_B)\\
 \alpha_{\pm}&=&J_x\mp J_y\end{array}.\nonumber\end{eqnarray}
where the matrix is the representation in the standard basis of
$\sigma_{Az}\sigma_{Bz}$ eigenstates. This state commutes with the spin parity
\cite{RCM.08} $P_z=-\exp[i\pi(\sigma_{Az}+\sigma_{Bz})/2]$. It is also
denoted as an $X$ state \cite{AR.10}.

We will now show that {\it a measurement of $\bm{\sigma}_{B}$ along  any of the
principal axes $x,y,z$ will provide a stationary point of $I_f^{\bm{k}}$
$\forall$ $S_f$}. \\
{\it Proof:} For a measurement along the $z$ axis ($\bm{k}=\bm{k}_z$), i.e.,
along the axis where $\rho_B$ is diagonal, $J^tJ\bm{k}_z=J_z^2\bm{k}_z$,
$\bm{r}_A$ and $\bm{r}_B$ are all along this  axis and Eq.\ (\ref{lam}) is then
trivially satisfied $\forall$ $\alpha_i$. It is a particular case of Eq.\
(\ref{d}), which here holds in the standard basis.

For a measurement along the $x$ axis ($\bm{k}=\bm{k}_x$),
$J^tJ\bm{k}_x=J_x^2\bm{k}_x$ while $\bm{r}_B\cdot \bm{k}_x=0$ and
$|\bm{r}_A+\nu J\bm{k}_x|=\sqrt{r_A^2+J_x^2}$. Hence
${p}^\mu_\nu=\frac{1}{4}(1+\mu|\bm{r}_A+\nu J\bm{k}_x|)$ is independent of
$\nu$. This leads to $\alpha_1=\alpha_2=0$ in (\ref{al}), in which case Eq.\
(\ref{lam}) is again satisfied. For $\bm{k}=\bm{k}_y$ the argument is similar.
We also remark that these arguments also apply to the quantum discord
(\ref{D}), as $\eta=0$ in (\ref{eqmod}) for $\bm{k}=\bm{k}_x$ or $\bm{k}_y$.

While other stationary directions may also exist, the principal axes are strong
candidates for minimizing $I_f^{\bm k}$. Typically, the minimum will be
attained for measurements along $z$ if ${\rm Max}[|J_x|,|J_y|]$ is sufficiently
small, while otherwise measurements along $x$ or $y$ will be preferred. A
transition between these two regimes will arise as $J_x$ or $J_y$ increases,
whose details will depend on the entropic function and may involve intermediate
directions $\bm{k}$.

Writing $\bm{k}=(\sin\gamma\cos\phi,\sin\gamma\sin\phi,\cos\gamma)$, these
intermediate solutions can be found from Eq.\ (\ref{lam}), which leads here to
$\phi=0$ or $\phi=\pi/2$ (if $|J_x|>|J_y|$ the minimum corresponds to $\phi=0$
for {\it any} $S_f$, as the ensuing distribution majorizes that for
$\phi=\pi/2$) and to $\gamma=0$ or
\begin{equation}
\cos\gamma=\frac{\alpha_1 r_{B}+\alpha_2 J_z r_A}
 {\alpha_3(J_x^2-J_z^2)}\,,\label{eqf}\end{equation}
where we have assumed $|J_x|>|J_y|$ such that $\phi=0$. The intermediate
solutions $|\gamma|\in(0,\pi/2)$ of (\ref{eqf}), if existent, are {\it
degenerate}, as both choices $\pm\gamma$ lead to the same $I_f^{\bm{k}}$. Just
the principal axes solutions are non-degenerate.

The final expression for $I_f^B$ is formally
\begin{equation}I_f^B(\rho_{AB})=\sum_{\mu,\nu=\pm}f({p}^\mu_\nu)-f(\lambda^\mu_\nu)\,,
 \label{Ifx}\end{equation}
where ${p}^\mu_\nu=\frac{1}{4}(1+\nu r_Bk_z+\mu\sqrt{(r_A+\nu
J_zk_z)^2+J^2_xk^2_x})$ are the eigenvalues (\ref{la}) of $\rho'_{AB}$ and
$\lambda^\mu_\nu$ those of $\rho_{AB}$:
\begin{equation}
\lambda^\mu_\nu={\textstyle\frac{1}{4}}[1+\nu J_z+
\mu\sqrt{(r_A+\nu r_B)^2+(J_x-\nu J_y)^2}]\,.
\label{pex}\end{equation}

We can verify the previous results in the quadratic and cubic cases. For an $X$
state both matrices $M_2$ and $M_3$ (Eqs.\ (\ref{M2}), (\ref{M3})) are diagonal
in the principal axes basis:
\begin{eqnarray}
M_{2_{\mu\nu}}&=&\delta_{\mu\nu}(J_\mu^2+\delta_{\mu z}r_B^2)\,,\nonumber\\
M_{3_{\mu\nu}}&=&\delta_{\mu\nu}[J_\mu^2+\delta_{\mu z}(r_B^2+2
r_B r_AJ _z)]\,.\nonumber
\end{eqnarray}
Hence, the optimum measurement will be along the axis with the maximum diagonal
value and {\it no intermediate solutions will arise} (for non-degenerate
eigenvalues), as opposed to the general case. Assuming $|J_y|<|J_x|$, a
``sharp'' $z\rightarrow x$ transition for the least disturbing  measurement
will then take place, the $x$ axis preferred for
 \begin{eqnarray} J_x^2&>&J_z^2+r_B^2\,,\;\;q=2\,,\label{c2}\\
 J_x^2&>&J_z^2+r_B^2+2r_Br_A J_z\,,\;\;q=3\label{c3}\end{eqnarray}
 in the quadratic and qubic cases respectively, such that
 \begin{eqnarray}
 I_2^B(\rho_{AB})&=&\half\{J_y^2+{\rm Min}[J_x^2,r_B^2+J_z^2]\}\,,\label{c4}\\
 I_3^B(\rho_{AB})&=&{\textstyle\frac{1}{4}}
 \{J_y^2-2J_xJ_yJ_z+{\rm Min}[J_x^2,r_B^2+J_z^2+2r_Ar_BJ_z]\}\,.\nonumber\\
 &&\label{c5}
 \end{eqnarray}
These expressions are in general no longer upper bounds to the squared
concurrence, which for these states is  $C_{AB}=\frac{1}{2}{\rm
Max}[|\alpha_+|-\sqrt{c_+c_-},|\alpha_-|-\sqrt{a_+a_-},0]$. Nonetheless,
$I_2^B$ remains an upper bound to $C_{AB}^2$ in the ``$z$ phase'', as
$C_{AB}^2\leq \frac{1}{4}(J_x\pm J_y)^2\leq
\frac{1}{2}(J_x^2+J_y^2)$.

\subsection{Mixture of aligned states}
As a particular relevant example of Eq.\ (\ref{X}), we will consider the
mixture of two states with spins aligned along different directions. Choosing
the $z$ axis as the bisector, such state can be written as
\begin{eqnarray}
\rho_{AB}&=&
\half(|\theta\theta\rangle\langle\theta\theta|
+|-\theta-\theta\rangle\langle-\theta-\theta|)\label{sth}\\
&=&
\frac{1}{4}\left(\begin{array}{cccc}a_+&0&0&c\\0&c&c&0\\
0&c&c&0\\c&0&0&a_-
\end{array}\right)\,,
\begin{array}{rcl}a_{\pm}&=&(1\pm\cos\theta)^2\\ c&=&\sin^2\theta
\end{array}\,,
\end{eqnarray}
which corresponds to $(J_x,J_y,J_z)=(\sin^2\theta,0,\cos^2\theta)$ and
$r_A=r_B=\cos\theta$ in (\ref{X}). Here
\begin{equation}{\textstyle|\theta\rangle=\exp[-i\frac{\theta}{2}\sigma_y]|0\rangle=
\cos\frac{\theta}{2}|0\rangle+\sin\frac{\theta}{2}|1\rangle}
\end{equation}
is the state with the spin forming an angle $\theta$ with the $z$ axis in the
$x,z$ plane. The relevance of this state was discussed in \cite{CRC.10}. It
represents, roughly,  the state of a spin pair in the definite parity ground
state of a finite $n$ spin ferromagnetic type $XY$ spin chain in a transverse
field for $|B|<B_c$, and the {\it exact} state of any pair at the immediate
vicinity of the factorizing field \cite{RCM.08} (neglecting small coherence
terms $\propto \cos^{n-2}\theta$).

This state is {\it separable}, i.e., it is a convex mixture of product states
\cite{RF.89}, and the concurrence $C_{AB}$ accordingly vanishes $\forall$
$\theta$. Nonetheless, it has non-zero discord \cite{CRC.10} if
$\theta\in(0,\pi/2)$. It will then have non-zero values of {\it any} $I_f^B$ in
this interval, with $I_f^B=I_f^A$ $\forall$ $S_f$ due to the symmetry of the
state. For $\theta=0$ it is obviously a pure product state, while for
$\theta=\pi/2$ it is a {\it classically correlated state}, i.e., diagonal in a
{\it standard} product basis, implying $I_f^B(\theta)\equiv
I_f^B(\rho_{AB}(\theta))=0$ for $\theta=0$ or $\theta=\pi/2$ $\forall$ $S_f$.

\begin{figure}[t]
\vspace*{-0.5cm}

\centerline{\scalebox{.6}{\includegraphics{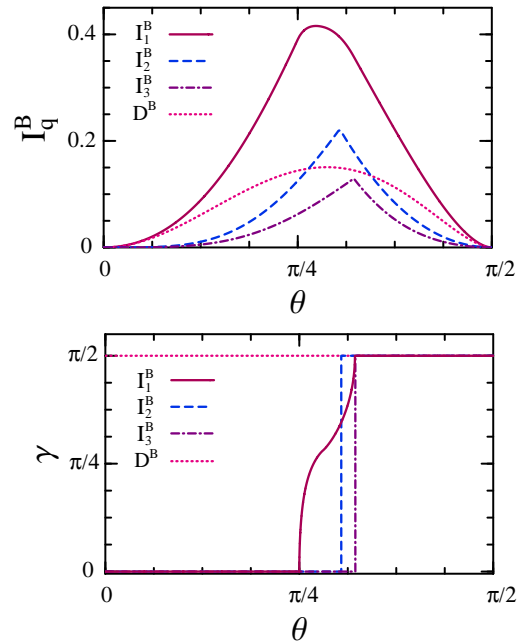}}}
 \vspace*{-0.5cm}

\caption{Top: The quantum correlation measures
$I_q^B(\rho_{AB})$ in the state (\ref{sth}),  as a function of the angle
$\theta$ for $q=1$ (von Neumann case), $2$  and $3$. $D^B$ denotes the quantum
discord. Bottom: The least disturbing measurement angle $\gamma$ vs.\ $\theta$
for the same cases depicted above. It is seen that $\gamma$ exhibits a sharp
transition from $0$ to $\pi/2$ (i.e., from $z$ to the $x$ axis) in the
quadratic $(q=2)$ and cubic $(q=3)$ cases, whereas in the von Neumann case
($q=1$) the transition is smooth. No transition arises in the case of the
quantum discord.}
 \label{f2}
\end{figure}
 \vspace*{0.cm}

It can be expected that as $\theta$ increases, the least disturbing measurement
will change from $z$ to $x$. In the quadratic and cubic cases, the transition
is {\it sharp}. We obtain, according to Eqs.\ (\ref{c2})--(\ref{c5}),
\begin{eqnarray}
I_2&=&\half\left\{\begin{array}{lr}
\sin^4\theta&\theta<\theta_{c2}\\ \cos^2\theta+\cos^4\theta
&\theta>\theta_{c2}\end{array},\right. \\
I_3&=&{\textstyle\frac{1}{4}}
\left\{\begin{array}{lr}\sin^4\theta&\theta<\theta_{c3}\\
\cos^2\theta+3\cos^4\theta &\theta>\theta_{c3}\end{array},\right.
 \end{eqnarray}
where $\cos^2\theta_{c2}=1/3$ ($\theta_{c2}\approx 0.61\pi/2$) and
$\cos^2\theta_{c3}=(\sqrt{17}-3)/4$ ($\theta_{c3}\approx 0.64\pi/2$), with the
minimizing measurement changing from $z$ to $x$ for $\theta>\theta_{ci}$. These
two quantities exhibit then a cusp-like maximum at $\theta=\theta_{ci}$, i.e.
slightly above $\pi/4$, as seen in Fig.\ \ref{f2}.

On the other hand, for other entropies a smooth transition from $z$ to the $x$
direction can arise. For instance, in the von Neumann case $z$ is preferred
exactly for $\theta\leq \pi/4$, but $x$ is minimum only for $\theta\agt
0.64\pi/2$. In between, the optimum measurement is obtained for an intermediate
angle $\gamma$, as determined by Eq.\ (\ref{eqf}), which varies continuously
from $0$ to $\pi/2$, as seen in Fig.\ \ref{f2}. This leads to a smooth maximum,
located closer to $\pi/4$. In the case of the quantum discord, the minimizing
angle is $\gamma=\pi/2$ $\forall$ $\theta$, exhibiting then {\it a different
behavior} due to the effect of the local term. In this case a relative entropy,
rather than a total entropy, is minimized.

\begin{figure}[t]
\vspace*{-0.5cm}

\centerline{\scalebox{.6}{\includegraphics{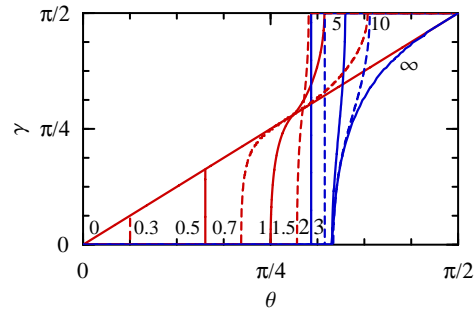}}}
 \vspace*{-0.5cm}

\caption{The least disturbing measurement angle $\gamma$ vs.\
$\theta$ determined by $I_q^B(\theta)$, for different values of $q$.}
 \label{f3}
\end{figure}
 \vspace*{0.cm}

For the present state there is no least mixed state $\rho'_{AB}$, and the least
disturbing measurement depends, therefore, on the entropic function. In order
to appreciate previous results from a more general perspective, the behavior of
the minimizing angle for different $q$ in the entropies (\ref{Sq}) is depicted
in Fig.\ \ref{f3}. The sharp transition $z\rightarrow x$ (i.e.,
$0\rightarrow\pi/2$) occurs for $2\leq q\leq 3$, indicating a special critical
behavior at these two values. A smoothed transition like that encountered in
the von Neumann case arises for $1/2<q<2$ and also $q>3$, where $\gamma$ varies
continuously from $0$ to $\pi/2$ within some window of $\theta$ values, which
narrows for $q$ close to $2$ or $3$.

For $0<q\leq 1/2$, the minimizing angle changes sharply from $0$ to an
intermediate value $\gamma\approx \theta$, increasing then almost linearly with
$\theta$ ($\gamma\approx\theta$). This is due to the fact that for low $q$,
$S_q(\rho'_{AB})$ is minimized when the lowest eigenvalue of $\rho'_{AB}$
vanishes, and this occurs precisely for $\gamma=\theta$. On the other hand, for
high $q$, $S_q(\rho'_{AB})$ is minimized when the largest eigenvalue of
$\rho'_{AB}$ is maximum, and the latter is maximized for $\gamma=0$ if
$\theta\leq\theta_c\approx 0.66\pi/2$, and for an intermediate $\gamma$ if
$\theta>\theta_c$, which varies continuously from $0$ to $\pi/2$ for
$\theta_c<\theta<\pi/2$. Accordingly, for high but finite $q$ values,
$\gamma=0$  for $\theta\alt \theta_c$, increasing then with $\theta$ and
reaching $\pi/2$ at an increasingly higher $\theta$. Different disorder
criteria lead then to different least disturbing measurements in this case, in
contrast with the state (\ref{rhox}).

\section{Conclusions\label{IV}}
We have analyzed the determination of the minimum information loss $I_f^B$
associated with an unread local measurement in a bipartite system, for a
general entropy $S_f$. Such quantity is a measure of the quantum correlations
lost in the local measurement, and reduces to the information deficit and the
geometric discord when $S_f$ is chosen as the von Neumann and linear entropy
respectively. A general stationary condition was derived, together with its
explicit form for an arbitrary mixed state of two qubits. Explicit expressions
for the cubic entropy and the associated measure $I_3^B$ were in this case
obtained, which require, as in the quadratic case (geometric discord), just the
eigenvalues of a $3\times 3$ matrix.

As application, we have first examined two-qubit mixed states with maximally
mixed marginals, where the minimum information loss $I_f^B$ for any entropy was
shown to be a simple function of the eigenvalues of $\rho_{AB}$. The minimizing
measurement is in this case universal. Moreover, in this case $I_2^B$ and
$I_3^B$ were shown to be strict upper bounds of the squared concurrence, which
is the associated entanglement monotone for both entropies. We have also
analyzed the case of $X$ states, providing explicit expressions for $I_2^B$ and
$I_3^B$ and showing that spin measurements along  the principal axes of the
matrix $J^tJ$ are {\it universal} stationary points of $I_f^B$ for {\it any}
$S_f$.

Finally, the special case of a mixture of aligned states was examined in
detail. Here the least disturbing local measurement changes, for all measures
$S_f$, from $z$ (bisector axis) to the $x$ axis as the angle $2\theta$ between
both directions changes from $0$ to $\pi$, being then different from that
optimizing the original quantum discord (which stays constant), although the
type of transition depends on the information measure employed. The least
disturbing measurement according to $I_f^B$ is thus more sensible to the
strength of the correlation, and reflects the ``transition'' experienced by the
state. Application of the present formalism to more complex systems is
currently under investigation.

The authors acknowledge support of CIC (RR) and CONICET (NC,LC) of Argentina.


\begin{thebibliography}{999}

\bibitem{BDSW.96} C.H.\ Bennett, D.P.\ DiVincenzo, J.A.\ Smolin, and W.K.\
Wootters, Phys. Rev. A {\bf 54}, 3824 (1996).
\bibitem{Be.93} C.H.\ Bennett et al., Phys.\ Rev.\ Lett.\ {\bf 70}, 1895
    (1993); Phys.\ Rev.\ Lett.\ {\bf 76}, 722 (1996).
\bibitem{NC.00}M.A.\ Nielsen and I.L.\ Chuang, {\it Quantum Computation and
               Quantum Information} (Cambridge Univ.\ Press, Cambridge, UK, 2000).
\bibitem{JL.03} R.\ Josza and N.\ Linden, Proc.\ R.\ Soc.\ {\bf A 459},
2011 (2003).
\bibitem{Vi.03} G.\ Vidal, Phys.\ Rev.\ Lett.\ {\bf 91}, 147902 (2003).
\bibitem{KL.98} E.\ Knill and R.\ Laflamme, Phys.\ Rev.\ Lett.\ {\bf 81},
5672 (1998).
\bibitem{DFC.05} A.\ Datta, S.T.\ Flammia and C.M.\ Caves, Phys.\ Rev.\ A
{\bf 72}, 042316 (2005).
\bibitem{RF.89} R.F.\ Werner, Phys.\ Rev.\ A {\bf 40}, 4277 (1989).
\bibitem{OZ.01} H.\ Ollivier and W.H.\ Zurek, Phys.\ Rev.\ Lett.\ {\bf 88},
 017901 (2001).
\bibitem{HV.01} L.\ Henderson and V.\ Vedral, J.\ Phys.\ A {\bf 34}, 6899 (2001).
\bibitem{Ve.03} V.\ Vedral, Phys.\ Rev.\ Lett.\ {\bf 90}, 050401 (2003).
\bibitem{ZZ.03} W.H.\ Zurek, Phys.\ Rev.\ A {\bf 67}, 012320 (2003).
\bibitem{DSC.08} A.\ Datta, A.\ Shaji, and C.M.\ Caves,
Phys.\ Rev.\ Lett.\ {\bf 100}, 050502 (2008).
\bibitem{Ho.05} M.\ Horodecki, et al.  Phys.\ Rev.\ A {\bf 71}, 062307
(2005); J.\ Oppenheim, M.\ Horodecki, P.\ Horodecki and R.\ Horodecki,
Phys.\ Rev.\ Lett.\ {\bf 89}, 180402 (2002).
\bibitem{SKD.11} A.\ Streltsov, H.\ Kampermann, and D.\ Bru\ss,
Phys.\ Rev.\ Lett.\ {\bf 106}, 160401 (2011).
\bibitem{DVB.10} B.\ Daki\'c, V.\ Vedral, and \v{C}.\ Brukner,
Phys.\ Rev.\ Lett.\ {\bf 105}, 190502 (2010).
\bibitem{RCC.10} R.\ Rossignoli, N.\ Canosa, and L. Ciliberti,
Phys.\ Rev.\ A {\bf 82}, 052342 (2010).
\bibitem{MV.10} K.\ Modi, et al, Phys.\ Rev.\ Lett.\ {\bf 104}, 080501 (2010).
\bibitem{SL.08} S.\ Luo, Phys.\ Rev.\ A {\bf 77}, 022301 (2008).
\bibitem{MD.11} V.\ Madhok and A.\ Datta, Phys.\ Rev.\ {\bf A} 83, 032323 (2011).
\bibitem{CAB.11} D.\ Cavalcanti et al, Phys.\ Rev.\ {\bf A} 83, 032324 (2011).
\bibitem{PGA.11} M.\ Piani et al, Phys.\ Rev.\ Lett.\ {\bf 106}, 220403 (2011).
\bibitem{RRV.11} L.\ Roa, J.C.\ Retamal, and M.\ Alid-Vaccarezza,
Phys.\ Rev.\ Lett.\ {\bf 107}, 080401 (2011).
\bibitem{ShL.09} A.\ Shabani and D.A.\ Lidar, Phys.\ Rev.\ Lett.\ {\bf 102}, 100402
(2009).
\bibitem{FA.10} A.\ Ferraro et al, Phys.\ Rev.\ A {\bf 81}, 052318 (2010).
\bibitem{FCOC.11}  F.F.\ Fanchini, M. F.\ Cornelio, M. C.\ de Oliveira, and A.O.\ Caldeira,
 Phys. Rev. A {\bf  84}, 012313 (2011).
\bibitem{LBAW.08} B.P.\ Lanyon, M.\ Barbieri, M.P.\ Almeida, A.G.\ White,
Phys.\ Rev.\ Lett.\ {\bf 101}, 200501 (2008).
\bibitem{DG.09} A.\ Datta and S.\ Gharibian, Phys.\ Rev.\ A {\bf 79}, 042325
(2009).
\bibitem{SL2.08} S.\ Luo, Phys.\ Rev.\ A {\bf 77}, 042303 (2008).
\bibitem{SA.09} M.S.\ Sarandy, Phys.\ Rev.\ A {\bf 80}, 022108 (2009).
\bibitem{WSF.09} T.\ Werlang, S.\ Souza, F. F.\ Fanchini,
and  C. J.\ Villas Boas, Phys.\ Rev.\ A {\bf 80}, 024103 (2009).
\bibitem{AR.10} M.\ Ali, A.R.P.\ Rau, and G. Alber, Phys.\ Rev.\ A {\bf  81}, 042105
(2010); ibid. {\bf 82}, 069902(E) (2010).
\bibitem{CRC.10} L. Ciliberti,  R.\ Rossignoli, and N.\ Canosa,
Phys.\ Rev.\ A {\bf 82}, 042316 (2010).
\bibitem{AD.11} D.\ Girolami and G.\ Adesso, Phys.\ Rev.\ A {\bf 83}, 052108 (2011).
\bibitem{YL.11} Y-C.\ Li and H-Q.\ Lin,  Phys.\ Rev.\ A {\bf 83}, 052323 (2011).
\bibitem{WW.98} W.K.\ Wootters, Phys.\ Rev.\ Lett.\ {\bf 80}, 2245 (1998).
\bibitem{Ca.03} P.\ Rungta and C.M.\  Caves, Phys.\ Rev.\  A {\bf 67},
012307 (2003); P.\ Rungta et al, Phys.\ Rev.\ A {\bf 64}, 042315  (2001).
\bibitem{RCM.08} R.\ Rossignoli, N.\ Canosa and J.M.\ Matera,
Phys.\ Rev.\ A {\bf 77}, 052322 (2008); R.\ Rossignoli, N.\ Canosa and J.M.\ Matera,
Phys.\ Rev.\ A {\bf 80}, 062325 (2009).
\bibitem{CR.02} N.\ Canosa and R.\ Rossignoli,  Phys.\ Rev.\ Lett.\ {\bf  88},
 170401 (2002).
\bibitem{WW.78} H.\ Wehrl, Rev.\ Mod.\ Phys.\ {\bf 50}, 221 (1978).
\bibitem{Bha.97} R.\ Bhatia, {\it Matrix Analysis} (Springer-Verlag, New York, 1997).
\bibitem{RC.03} R.\ Rossignoli, N.\ Canosa, Phys.\ Rev.\ A {\bf 67}, 042302
(2003). R.\ Rossignoli, N.\ Canosa, Phys.\ Rev.\ A {\bf 66}, 042306 (2002).
\bibitem{Ve.02} V.\ Vedral, Rev.\ Mod.\ Phys.\ {\bf 74}, 197 (2002).
\bibitem{TS.88}C.\ Tsallis, J. Stat. Phys. {\bf 52}, 479 (1988);
C.\ Tsallis, {\it Introduction to non-extensive statistical mechanics}
(Springer, New York, 2009).
\bibitem{Vi.00} G.\ Vidal, J.\ Mod.\ Opt.\ {\bf 47}, 355 (2000).
\end{thebibliography}
\end{document}